\documentclass[aps,twocolumn,groupedaddress,floatfix]{revtex4-1}
\usepackage{amssymb,amsmath}
\usepackage{graphicx}
\usepackage{subfigure}
\usepackage[english]{babel}
\usepackage{float}
\usepackage{color}

\usepackage{lipsum}
\usepackage{suffix}
\usepackage{mathtools}

\DeclarePairedDelimiterX\MeijerM[3]{\lparen}{\rparen}%
{\begin{smallmatrix}#1 \\ #2\end{smallmatrix}\delimsize\vert\,#3}

\newcommand\MeijerG[8][]{%
  G^{\,#2,#3}_{#4,#5}\MeijerM[#1]{#6}{#7}{#8}}

\WithSuffix\newcommand\MeijerG*[7]{%
  G^{\,#1,#2}_{#3,#4}\MeijerM*{#5}{#6}{#7}}
\begin{document}
\newcommand{\be}{\begin{equation}}
\newcommand{\ee}{\end{equation}}
\newcommand{\rojo}[1]{\textcolor{red}{#1}}

\title{Magnetic metamaterials with correlated disorder}

\author{Mario I. Molina}
\affiliation{Departamento de F\'{\i}sica, Facultad de Ciencias, Universidad de Chile, Casilla 653, Santiago, Chile}

\date{\today }

\begin{abstract} 
We examine the transport of magnetic energy in a simplified model of a magnetic metamaterial, consisting of 
a one-dimensional array of split-ring resonators, in the presence of correlated disorder in the resonant frequencies. The computation of the average participation ratio (PR) reveals that on average, the modes for the correlated disorder system are less localized than in the uncorrelated case. The numerical computation of the mean square displacement of an initially localized magnetic excitation for the correlated case, shows a substantial departure from the uncorrelated (Anderson-like) case. A long-time asymptotic fit $\langle n^2 \rangle \sim t^{\alpha}$ reveals that, for the uncorrelated system $\alpha\sim 0$, while for the correlated case $\alpha>0$, spanning a whole range of behavior ranging from localization to super-diffusive behavior. The transmission coefficient of a plane wave across a single magnetic dimer reveals the existence of well-defined regions in disorder strength-magnetic coupling space, where unit transmission for some wavevector(s) is possible. This implies, according to the random dimer model (RDM) of Dunlap et al., a degree of mobility. A comparison between the mobilities of the correlated SRR system and the RDM shows that the RDM model has better mobility at low disorder while our correlated SRR model displays better mobility at medium and large disorder.

\end{abstract}

\maketitle

{\em Introduction}. The subject of metamaterials has continue to attract interest in the scientific community, for its great potential for applications to many different technologies.
We can briefly describe them as a class of man-made materials that are characterized by having enhanced thermal, optical, and transport properties that make them attractive candidates for current and future technologies. Among them, we have magnetic metamaterials (MMs) that consist of artificial structures whose magnetic response can be tailored to a certain extent. A simple realization of such a system consists of an array of metallic split-ring resonators (SRRs) coupled inductively\cite{SRR1, SRR2, SRR3}. This type of system can, for instance, feature negative magnetic response in some frequency window, making them attractive for use as a constituent in negative refraction index materials\cite{pendry,veselago,negative_refraction,padilla}. The main drawback of this system is the existence of large ohmmic and radiative losses. A possible way out that has been considered is to endow the SRRs with external gain, such as tunnel (Esaki) diodes\cite{losses1,losses2} to compensate for such losses. The theoretical treatment of such structures relies mainly on the effective-medium approximation where the composite is treated as a homogeneous and isotropic medium, characterized by effective macroscopic parameters. The approach is valid, as long as the wavelength of the electromagnetic field is much larger than the linear dimensions of the MM constituents.

The simplest MM model utilizes an array of split-ring resonators (Fig.1), 
with each
resonator consisting of a small, conducting ring with a
slit. Each SRR unit in the array can be mapped to a
resistor-inductor-capacitor (RLC) circuit featuring self-
inductance $L$, ohmic resistance $R$, and capacitance $C$
built across the slit. We will assume a negligible resistance, 
and thus each unit will possess a resonant frequency $\omega=1/\sqrt{L C}$.
\begin{figure}[t]
 \includegraphics[scale=0.125]{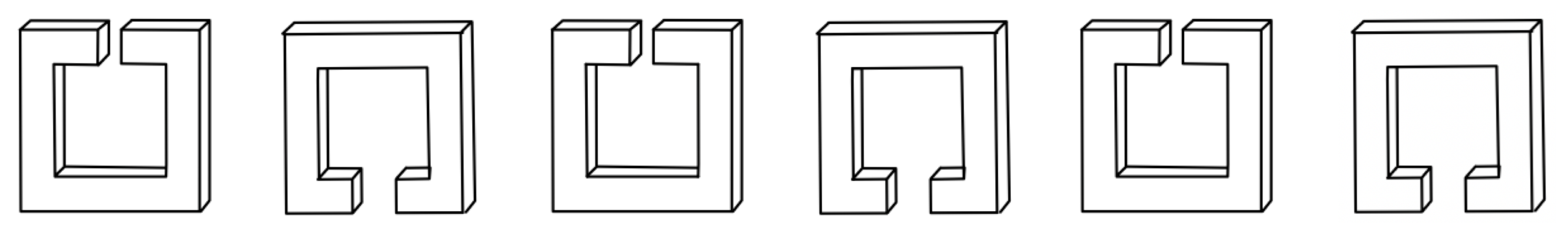}\\
 \includegraphics[scale=0.125]{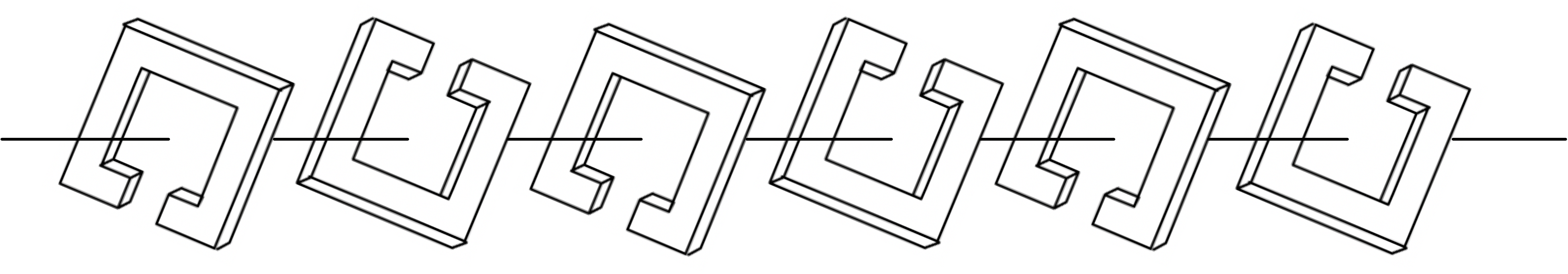}
 
  \caption{One-dimensional split-ring resonator arrays. Top: All SRRs lying on a common plane ($\lambda < 0$). Bottom: All SRRs parallel but centered around a common axis ($\lambda > 0$)}
  \label{fig1}
\end{figure}
Under this condition and in the absence of driving and dissipation, the dimensionless evolution equations for the charge $q_{n}$ residing at the $nth$ ring are
\be
{d^2\over{d t^2}}\left( q_{n} + \lambda (q_{n+1} + q_{n-1})\right) + \omega_{n}^{2} q_{n} = 0 \label{eq1}
\ee
where $q_{n}$ is the dimensionless charge of the nth ring,  $\lambda$ is the coupling between neighboring rings which originates from the dipole-dipole interaction, and $\omega_{n}^2$ is the resonant frequency of the $nth$ ring, normalized to a characteristic frequency of the system, such as the average frequency: $\langle w_{n}^2\rangle=(1/N) \sum_{n} w_{n}^2$. The frequency $\omega_{n}^2$ can be changed by varying the capacitance of the ring by altering the slit width or by inserting a dielectric in the slit. For a homogeous array, $\omega_{n}^2=1$.

The dimensionless stationary state equation is obtained from Eq.(\ref{eq1}) after
posing $q_{n}(t) = q_{n} \exp[i (\Omega t + \phi) ]$:
\be
-\Omega^2 \left( q_{n} + \lambda (q_{n+1} + q_{n-1})\right) + \omega_{n}^{2} q_{n}=0
\label{eq2}
\ee
On the other hand, the topic of Anderson localization is an old one, but its importance has no waned throughout  the years, given its consequences for transport in disordered systems.  Roughly speaking, it asserts that the presence of disorder tends to inhibit the propagation of excitations. In fact, for 1D systems, all the eigenstates are localized and transport is completely inhibited\cite{anderson1, anderson2, anderson3}. Now, Anderson localization is based on the assumption that the disorder is ``perfect'' or uncorrelated. However it has been noted that in systems with correlated disorder, a degree of transport is still possible. Such is the case of the random dimer model (RDM) for the discrete Schr\"{o}dinger equation, that consists on a binary alloy where one of the site energies is assigned at random to pairs of lattice sites. This leads to a mean square displacement of an initially localized excitation that grows asymptotically as $t^{3/2}$ at low disorder levels, instead of the saturation behavior predicted by Anderson theory\cite{correlated1,correlated2,correlated3}. An experimental demonstration of the RDM prediction has been made in an optical setting\cite{szameit}. A straightforward extension of these ideas to random arrays of larger units (n-mers), has also been theoretically explored\cite{nmers}. Magnetic metamaterials constitute yet another setting in which to test all these ideas, whose results could have an impact on the design of future materials of technological importance.

In this work, we will explore the stationary modes and the transport of excitations in the context of a magnetic random dimer model. As we will see, the presence of uncorrelated disorder leads to fully localized modes and to the absence of transport. On the contrary, the presence of a short-range correlation in the disorder distribution leads to an improved transport of magnetic energy, in qualitative agreement with the standard RDM. The results of these studies give us some inkling as to the magnetic energy transport in a correlated  disordered SRR  array, as well as to checking the universality of Anderson localization. 

The `size' or extent of the distribution of electric charge $\{q_{n}(t)\}$ stored in the capacitors, can be monitored via the participation ratio (PR), defined as
\be
PR = (\ \sum_{n} |q_{n}(t)|^2 \ )^2/\sum_{n} |q_{n}(t)|^4
\ee
For a completely localized excitation, $PR=1$, while for a complete delocalized state, 
$PR=N$.

To monitor the degree of mobility of a magnetic excitation propagating inside a SRR array, we resort to the mean square displacement (MSD) of the charge distribution, defined as
\be
\langle n^2 \rangle = \sum_{n} n^2 |q_{n}(t)|^2 / \sum_{n} |q_{n}(t)|^2\label{n2}.
\ee
Typically $\langle n^2 \rangle\sim t^\alpha$ at large $t$, where $\alpha$ is known as the transport exponent. The types of motion are classified according to the value of $\alpha$: `localized' ($\alpha=0$), `sub-diffusive' ($0<\alpha<1$), `diffusive' ($\alpha=1$), `super-diffusive' ($1<\alpha<2$) and `ballistic' ($\alpha=2$).

{\em Homogeneous case}. In the absence of disorder, $\omega_{n}^2 =  1$, and the discrete translational invariance leads to an energy band, obtained from Eq.(\ref{eq2}) after assuming a plane wave profile $q_{n} \sim \exp(i k n)$:
\be 
\Omega_{k}^2 = {1\over{1 + 2 \lambda \cos(k)}}\label{eq3}
\ee
This implies that these magneto-inductive waves can only exist for $|\lambda|<1/2$. From Lenz law, it can be shown that when all SRRs lie on a common plane, $\lambda<0$, while when all SRRs are centered about a common axis, $\lambda>0$ (Fig.\ref{fig1}).

For a completely localized initial charge $q_{n}(0)=A\ \delta_{n 0}$ and no currents, $(d q_{n}/d t)(0)=0$, we have formally
\be q_{n}(t) = (A/4 \pi) \int_{-\pi}^{\pi} e^{i (k n-\Omega_{k} t)} dk + (A/4 \pi) \int_{-\pi}^{\pi} 
e^{i (k n+\Omega_{k} t)} dk
\ee
where $\Omega_{k}$ is given by Eq.(\ref{eq3}). After replacing this form for $q_{n}(t)$ into Eq.(\ref{n2}), one obtains after some algebra, a closed form expression for $\langle n^2 \rangle$:
\be
\langle n^2 \rangle = {(1/2 \pi)\int_{-\pi}^{\pi}d k (d \Omega_{k}/d k)^2 (1 - \cos(2\ \Omega_{k}\ t))\ t^2 
\over{1 + (1/2\pi) \int_{-\pi}^{\pi} d k\ \cos(2\ \Omega_{k}\ t)}}\label{n2closed}
\ee
As we can see from the structure of Eq.(\ref{n2closed}), as time $t$ increases, the contributions from the cosine terms to the integrals decrease and, at long times, $\langle n^2\rangle$ approaches a ballistic behavior
\be
\langle n^2 \rangle =\left[ {1\over{2 \pi}} \int_{-\pi}^{\pi} \left( {d \Omega(k)\over{d k}}\right)^2\ dk\right]\ t^2\hspace{1cm} (t\rightarrow \infty),\label{RMS}
\ee
while at short times,
\be
\langle n^2 \rangle = \left[ {1\over{2 \pi}} \int_{\pi}^{\pi} \left( \Omega_{k}{d \Omega_{k}\over{d k}} \right)^2 dk\right] \ t^4 \hspace{1cm}(t\rightarrow 0).
\ee
Thus, the transport in our system is ballistic: $\langle n^2 \rangle =g(\lambda)\ t^2$, where we can identify $\sqrt{g(\lambda)}$ as a kind of characteristic `speed' for the ballistic propagation. Inserting the specific form for $\Omega_{k}$ from Eq.(\ref{eq3}),
we obtain
\be
\langle n^2 \rangle = {(\lambda t)^2\over{2 (1 - 4 \lambda^2)^{3/2}}}\hspace{1cm} t\rightarrow \infty ,
\ee
\be
\langle n^2 \rangle = {\lambda^2 (1+\lambda^2) t^4\over{2 (1 - 4 \lambda^2)^{3/2}}}\hspace{1cm} t\rightarrow 0.
\ee

{\em Disordered case}. In this case, we choose the resonant frequencies $\omega_{n}^2$ from a random binary distribution $\{\omega_{a}^2,\omega_{b}^2\}$. As mentioned before, this can be achieved by altering the space between the slits of each SRR, or by filling the space in the slit with different dielectrics. Let us look at the stationary modes obtained from Eq.(\ref{eq2}). After a simple rearrangement, one can write down an equivalent  eigenvalue problem:
\be
-\left({1\over{\Omega^2}} \right) q_{n} + \left({1\over{\omega_{n}^2}}\right) q_{n} + \lambda \left({1\over{\omega_{n}^2}}\right) (q_{n+1} + q_{n-1}) = 0
\ee
This looks similar to the tight-binding Anderson problem, except that now, we have a completely correlated site and coupling randomness. Not only that, but the site energies and couplings appear all `inverted'. This means that the usual, large disorder limit of the tight-binding system corresponds here to the small disorder limit. 

The rigid correlation between the site `energies' and the coupling terms will be of lesser importance than the other correlation we really have in mind:  that of a random binary alloy for the resonance frequencies, as in the RDM. Following the RDM, we assign the site frequencies $\omega_{n}^2$ at random to {\bf pairs} of lattice sites (that is, two sites in succession): $...\omega_{a}^2,\omega_{a}^2,\omega_{b}^2,\omega_{b}^2,\omega_{b}^2,\omega_{b}^2,\omega_{a}^2,\omega_{a}^2,\omega_{b}^2,\omega_{b}^2,\omega_{a}^2,\omega_{a}^2,...$. The frequency of each pair is generated according to $\omega_{n}^2 = \omega_{a}^2 + (\omega_{b}^2-\omega_{a}^2)\times \mbox{rand}$, where rand=$0$ or $1$ with fifty percent probability. It should be emphasized that our system cannot be mapped to the original RDM because of the the additional correlation between site and coupling values.

In what follows, we will examine the stationary and transport properties of a SRR array with binary disorder (``correlated'') and that of a simple Anderson-like (``uncorrelated'') disordered SRR array. 

\begin{figure}[t]
 \includegraphics[scale=0.3]{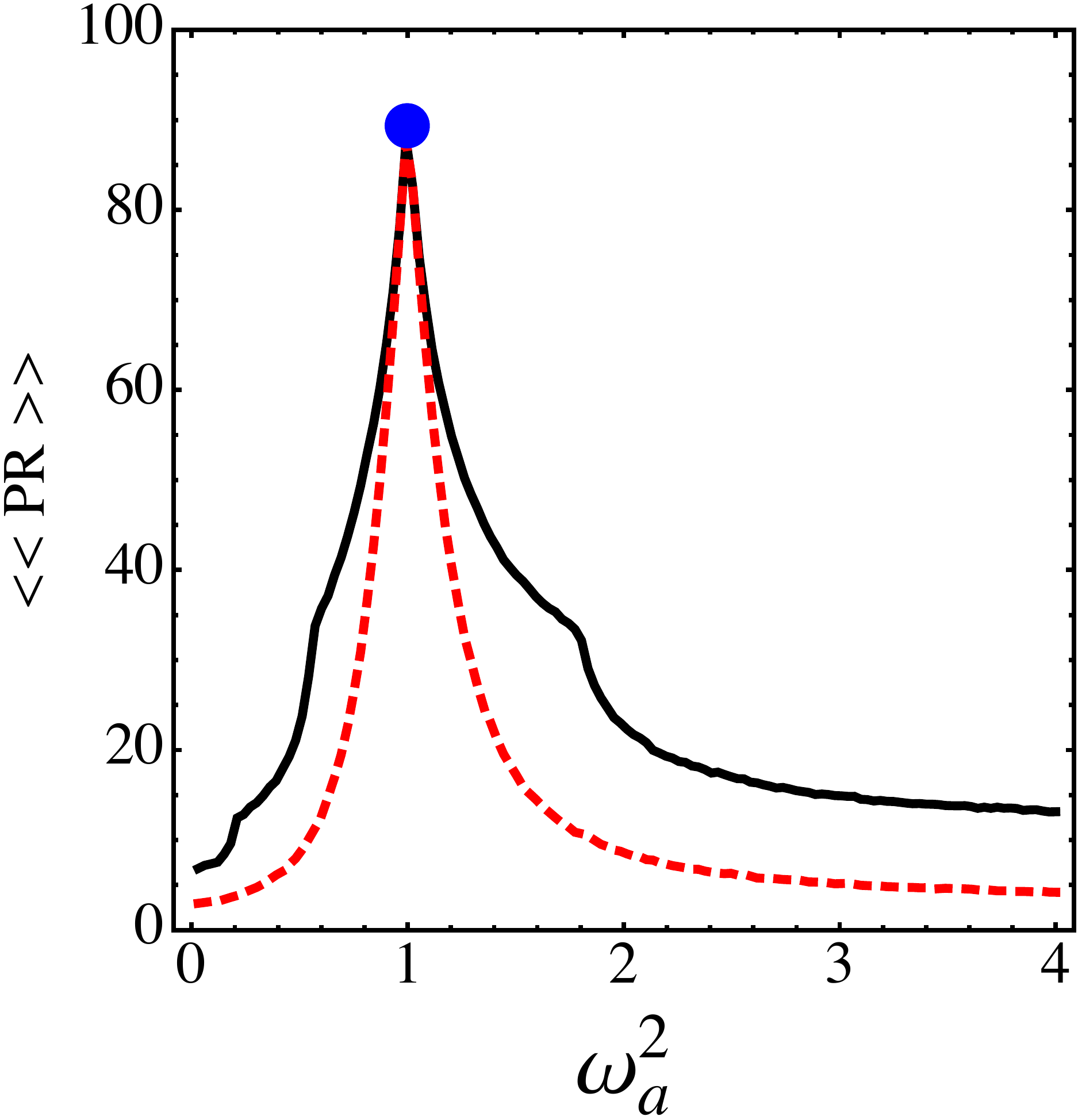}
  \caption{Mode -and realization average participation ratio for the correlated SRR (solid line) and uncorrelated SRR case (dashed line), as a function of disorder strength. The dot marks the value of the participation ratio in the absence of disorder ($N=150, \lambda=0.4, \omega_{b}^2=1, \mbox{realizations}=100$.)}
  \label{fig2}
\end{figure}

Figure \ref{fig2} shows the mode-and realization average of the participation ratio, as a function of disorder strength. As we can see, in both cases, the presence of disorder makes the PR smaller than in the periodic case  $\omega_{a}=\omega_{b}$, where $PR=(2/3)N$. This last case corresponds in our plot to $\omega_{a}^2=1$. In the correlated case, the PR is alway greater than in the uncorrelated case, which means that the modes are more extended in space. This, in turn, means that there is more overlap between modes which leads to an increased mode coupling. An increased mode coupling means greater ease for an excitation to jump to nearby sites, thus increasing the general mobility.

A qualitative explanation for this enhanced transport can be given within the context of the RDM. The idea is that the presence of the correlated binary disorder creates a finite fraction of modes with localization length equal or greater to the length of the array. Adapting the RDM argument, we consider an array of SRRs with identical resonance frequencies $\omega_{n}=1$, that contains an embedded dimer impurity. Without loss of generality, we place the dimer at $n=0$ and $n=1$ and set $\omega_{0}=\omega_{1}=\omega$. Let us consider an incoming  plane wave and compute its transmission coefficient across the dimer. The amplitudes to the left and right of the dimer impurity are
\be
q_{n}=\left\{ 
				\begin{array}{ll}
				R_{0} \exp(i k n) + R \exp(-i k n)\hspace{0.5cm}n\leq 0\\
					T \exp(i k n)\hspace{3cm}n\geq 1
					\end{array}
					\right.
\ee
where $R_{0}$ is the incoming amplitude, $R$ the reflected amplitude and $T$ the transmission amplitude. The stationary equations at the dimer sites are
\begin{eqnarray}
\left[-\left({1\over{\Omega^2}} \right) + \left({1\over{\omega^2}} \right)\right](R_{0}+R)+ & &\\\left({\lambda\over{\omega^2}}\right)(T e^{i k}+ R_{0} e^{-i k}+ R e^{i k})&=&0\ \ \ \ \ \ \ \\
\left[-\left({1\over{\Omega^2}} \right) + \left({1\over{\omega^2}} \right)\right] T e^{i k} + & & \\
+\left({\lambda\over{\omega^2}}\right) (T e^{2 i k}+R_{0}+R)& = &0\ \ \ \ \ \ \ 
\end{eqnarray}
together with $\Omega^2 = 1/(1 + 2 \lambda \cos(k))$.

From these equations, it is possible to find the transmission coefficient $t(k)=|T(k)|^2/|R_{0}|^2$ in closed form as
\begin{widetext}
\be
t(k) = \left| (-1+e^{2 i k}) \lambda^2\over{2 e^{3 i k}\lambda(\omega^2 -1)+e^{4 i k}\lambda^2(\omega^2 -1)+2 e^{i k}\lambda(\omega^2 -1)\omega^2+\lambda^2\omega^4+e^{2 i k}( (\omega^2 -1)^2+\lambda^2(-1+2(\omega^2 -1)\omega^2) )}  \right|^2.\label{eq17}
\ee
\end{widetext}
In the absence of the dimer, $\omega^2=1$ and $t(k)=1$. 
\begin{figure}[t]
 \includegraphics[scale=0.3]{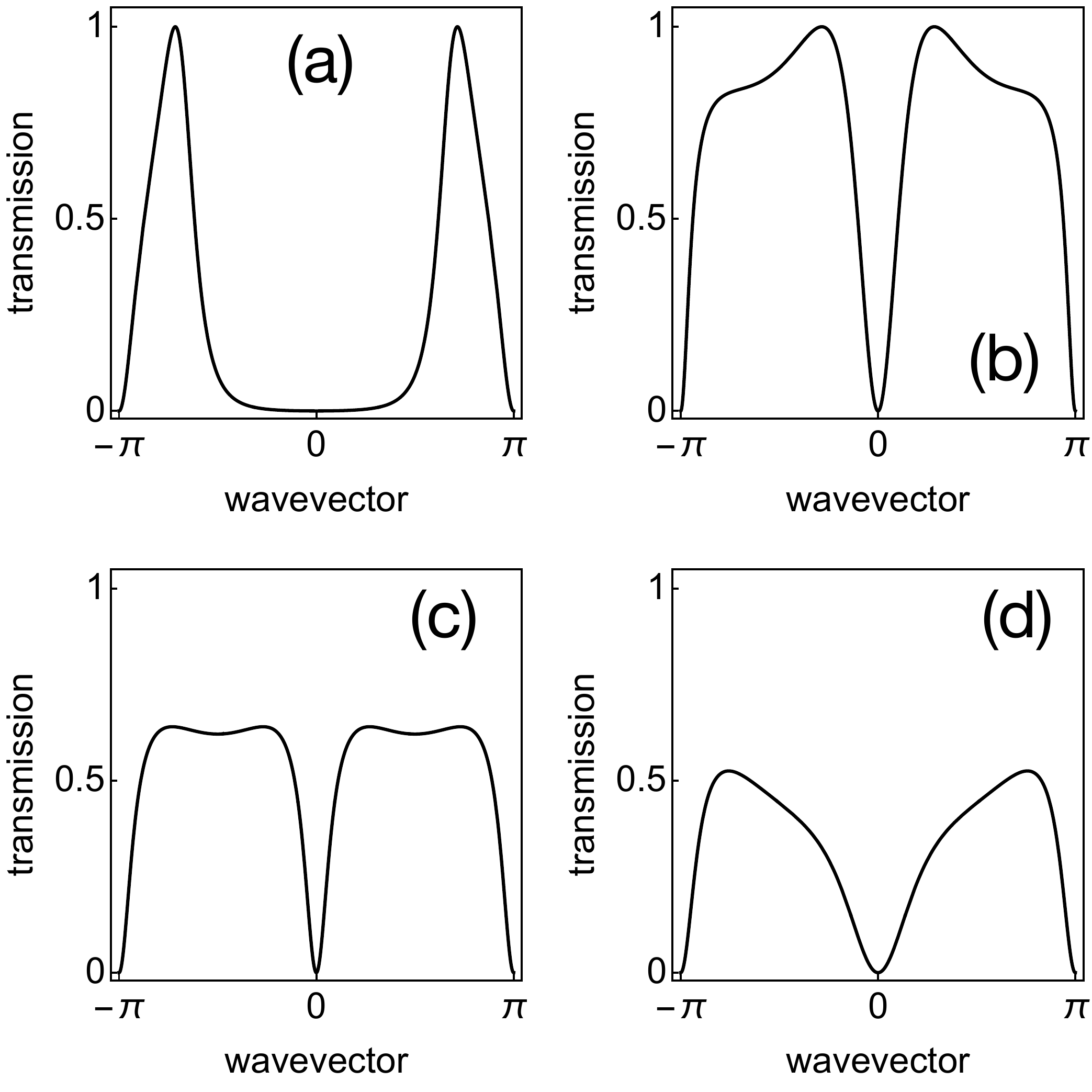}
  \caption{Transmission coefficient of magneto-inductive plane waves across a SRR dimer, for different values of the resonant frequency mismatch. (a) $\omega^2=2$, (b) $\omega^2=2/3$, (c) $\omega^2=0.5$, (d) $\omega^2=0.4$ ($\lambda=0.4$).}
  \label{fig3}
\end{figure}
\noindent
Figure \ref{fig3} shows some transmission plots for a fixed value of $\lambda$ and different $\omega^2$. The most interesting thing to notice is the existence of resonant cases. That is, the existence of wavevectors $k$ at which $t(k)=1$, for some values of $\lambda, \omega^2$. To determine the resonant region in parameter space  we make a sweep of $t(k)$ in $\lambda, k$ and $\omega^2$. Results are shown in Fig.\ref{fig4} that shows regions of possible resonance-no resonance in parameter space.
\begin{figure}[t]
 \includegraphics[scale=0.375]{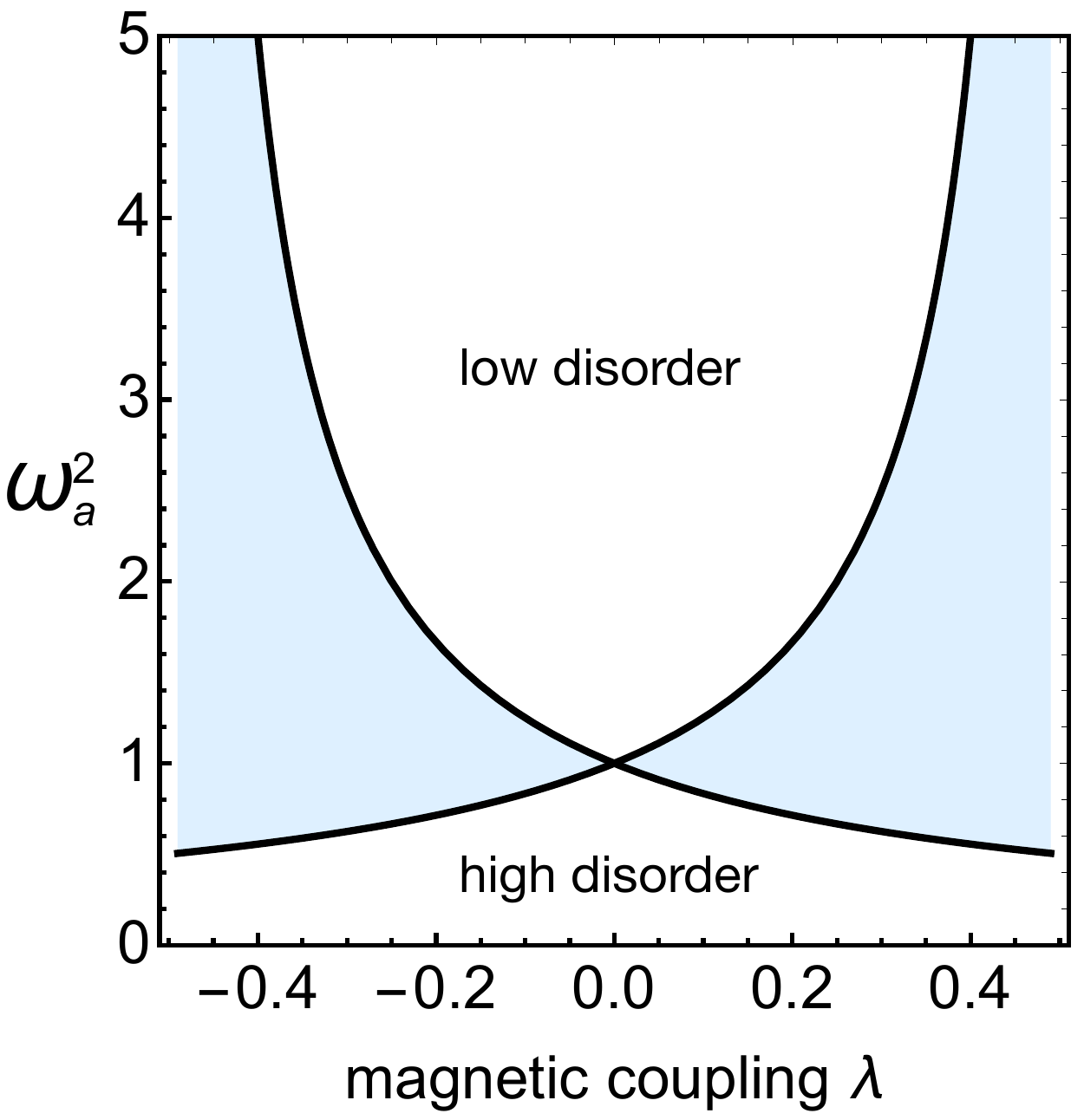}
  \caption{Region in parameter space where a resonant transmission(s) across the SRR dimer is possible (shaded region). ($\omega_{b}^2=1$)}
  \label{fig4}
\end{figure}
This region corresponds to the area enclosed by the curves $1/(1+ 2 \lambda)$ and $1/(1 - 2 \lambda)$.
Inside the region, a resonance(s) across a single dimer is possible, and we will obtain an enhanced transport behavior when we form a random dimer alloy: Once a plane wave goes through a dimer without reflection, it will pass through all of the other SRRs dimers and will reach the ends of the system unscattered. Of course this only happens for a single wavevector, so it might be regarded as a marginal effect. However, around the perfectly transmitting case, there will be a fraction of states whose transmission across the system is finite\cite{correlated2} and, in our case,  will give rise to some degree of magnetic transport.

\begin{figure}[t]
\hspace{1cm}\includegraphics[scale=0.25]{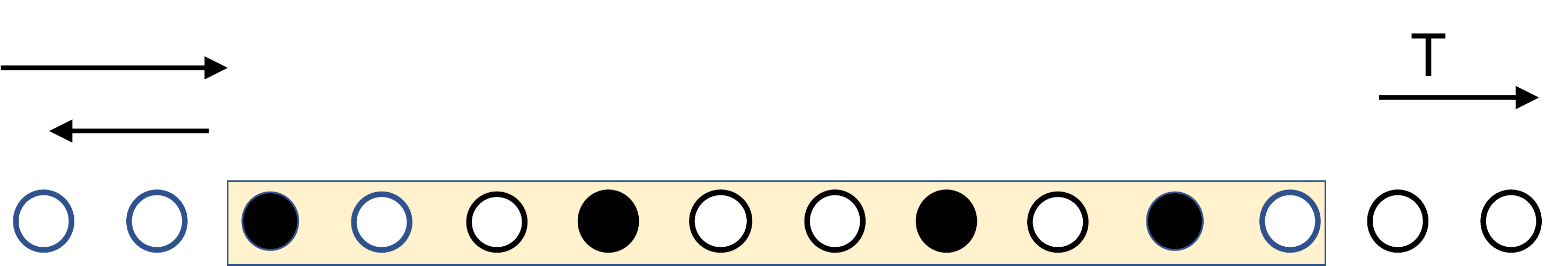}\\
 \includegraphics[scale=0.275]{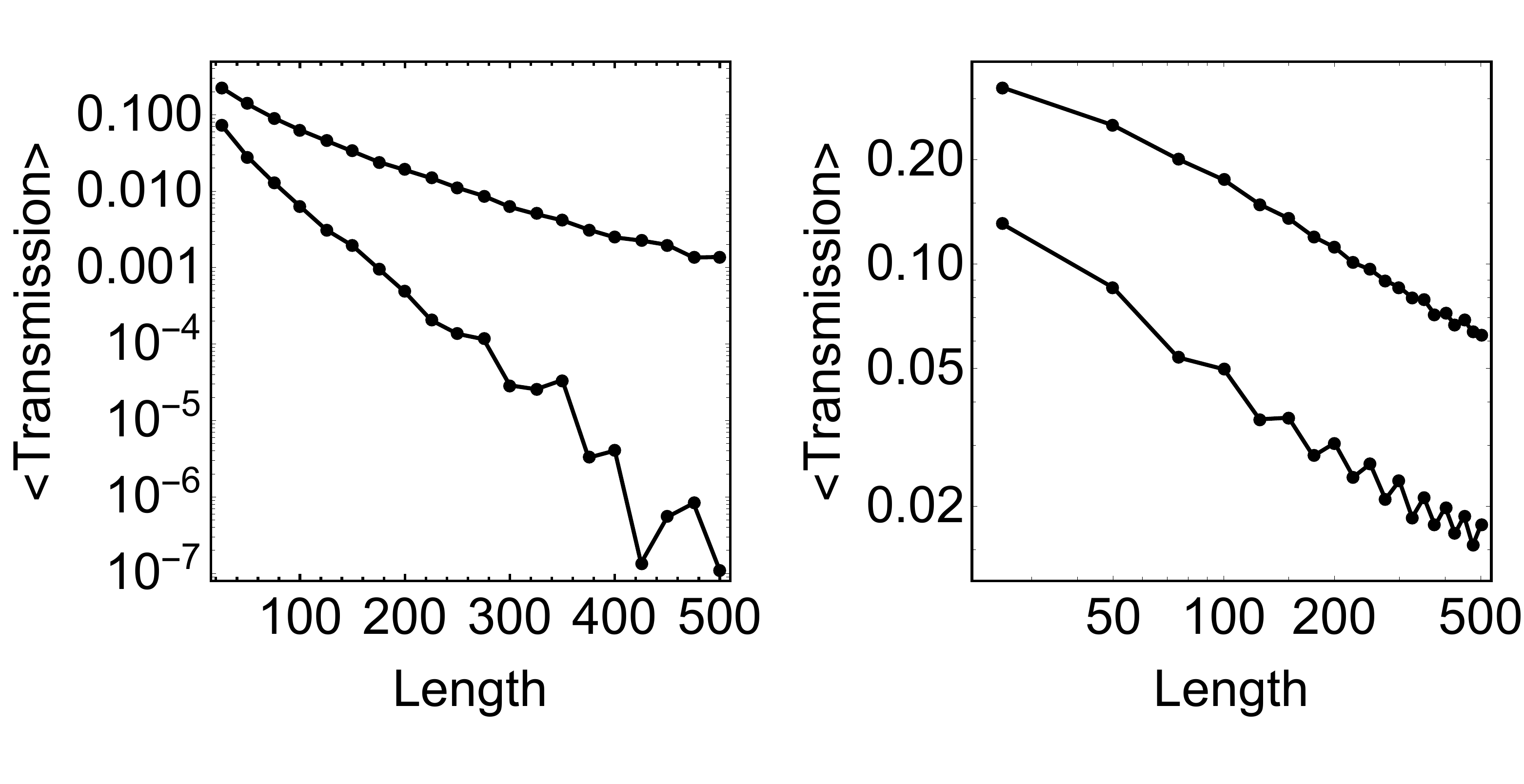}
  \caption{wavevector -and realization average transmission of plane waves across a weakly disordered segment of finite length. (a) Uncorrelated case (b) Correlated case. Note the logarithmic scale in (b). Upper curves: $\omega_{a}^2=1.5, \omega_{b}^2=1$. Lower curves: $\omega_{a}^2=2, \omega_{b}^2=1$ ($\lambda=0.4, \mbox{realizations=500}$).}
  \label{fig405}
\end{figure}

Let us also compute the average transmission of a plane wave across an extended portion of the disordered array. Results are shown in Fig.\ref{fig405}. For weak disorder ($\omega_{a}^2 \gg 1, \omega_{b}^2=1$) we see that for the uncorrelated case the transmission decreases 
exponentially with system size $\langle T\rangle \sim \exp(-\alpha L)$, while for the correlated case the decrease obeys a power-law $\langle T\rangle \sim L^{-\beta}$. These results are in qualitative agreement with previous studies on tight-binding systems\cite{nmers}. For large disorder strengths ($\omega_{a}^2 \ll 1, \omega_{b}^2=1$), we obtain exponential decrease for both cases (not shown).

\begin{figure}[t]
\includegraphics[scale=0.2]{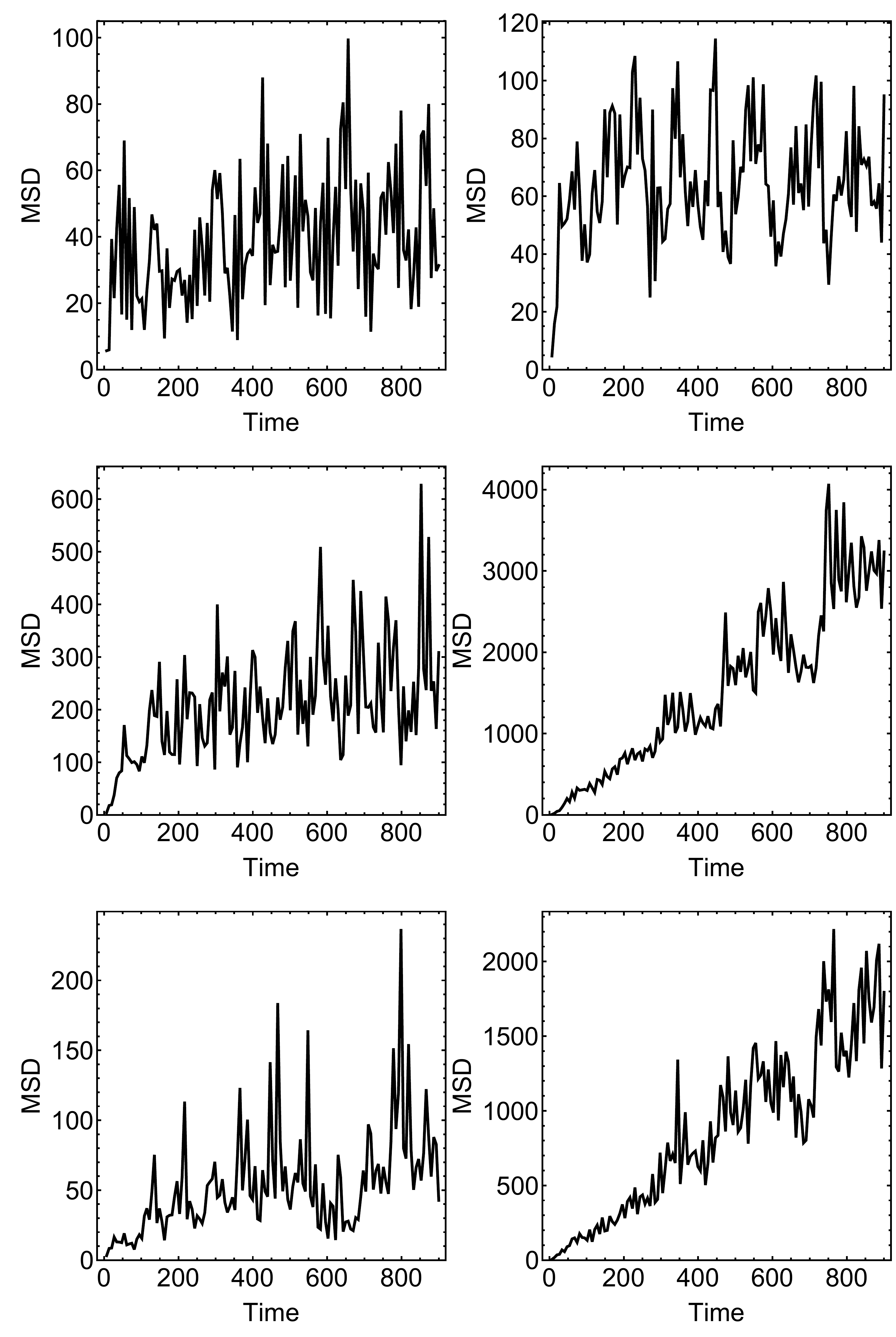}
  \caption{Mean square displacement as a function of time, for a SRR array with a localized initial condition and different disorder widths, for both, uncorrelated (left column) and correlated (right column) cases.
 (a) and (b): $\omega^2_{a}=0.1 \lambda$. (c) and (d): $\omega_{a}^2=2 \lambda$. (e) and (f): $\omega^2_{a}=3 \lambda$ $(\lambda=0.4, \omega_{b}^2=1, N=1500, t_{max}=900$).}
  \label{fig5}
\end{figure}

\begin{figure}[t]
 \includegraphics[scale=0.35]{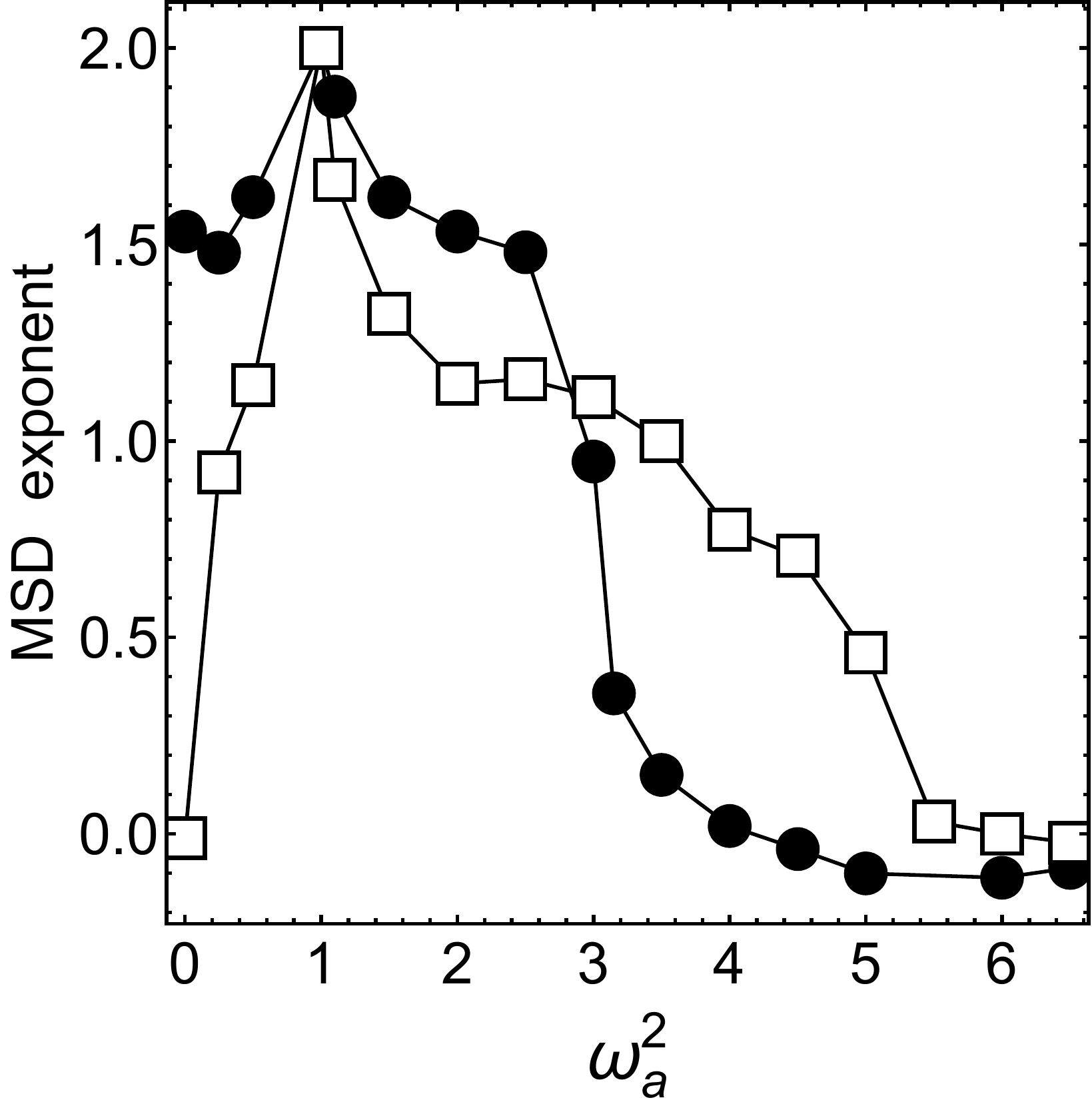}
  \caption{Mean square displacement exponent $\alpha$ as a function of disorder strength, for  an initially localized magnetic excitation. Squares (circles) denote the SRR correlated (RDM correlated) case. ($\lambda=0.4, \omega_{b}^2=1, N=1500, t_{max}=900$).}
  \label{fig6}
\end{figure}

{\em Transport}. Let us compute the mean square displacement (MSD) of an initially localized magnetic excitation, and compare the results for the correlated and uncorrelated cases.
Results are shown in figure \ref{fig5}. 
In this figure we compare the MSD for the uncorrelated case (left) column with the correlated one (right column), for three common values of the disorder strength $\omega_{a}^2$. The plots are computed for a single realization of a long SRR array and long evolution times. For all of them we use the same skeleton random number sequence, and vary only $\omega_{a}^2$ while keeping $\omega_{b}^2=1$. In this way, the value of $\omega_{a}^2$ controls the disorder width. As we can see, while for the uncorrelated case the MSD tends to saturate at long times, in the correlated case there is a  finite fraction of propagation at long times. Not only that, but the $\omega_{a}^2$ values separating mobile from localized regions are in agreement with the 
mobility phase diagram of Fig. \ref{fig4}. Assuming a long-time asymptotic dependence of the form $\langle n^2 \rangle \sim t^\alpha$, we computed numerically  the $\alpha$ values as a function of the disorder strength $\omega_{a}^2$. Results are shown in Fig.\ref{fig6}, and are compared to the tight-binding RDM model of Dunlap et al.\cite{correlated2}.
Interestingly, at lower disorder strength the RDM shows higher exponents (i.e., better propagation). In this range, the exponent values are close to the superdiffusive value: $\alpha=3/2$. For the correlated SRR  case, the propagation is closer to the diffusive regime: $\alpha=1/2$. At intermediate and high disorder strengths  however, this tendency reverts and the MSD for the SRR model shows substantially higher exponents, signaling an enhanced mobility. In the limit of large disorder, both sets of exponents converge to zero. Of course, these observations are only valid for a finite array; the real asymptotic exponents are defined for an infinite  system only.  

{\em Conclusions}.
In this work we have examined the stationary modes and the propagation characteristics of a SRR disordered array. We have employed two types of disorder: An Anderson-like with randomly distributed resonant frequencies, and a correlated one, where the frequencies distributions follows the RDM pattern, that is, a disordered binary 
distribution. In the absence of disorder, the mean square displacement is calculated in closed form, and at long times shows a ballistic behavior. In the presence of disorder, the stationary modes for a finite array are mostly localized, with a small fraction possessing a localization length of the size of the system's length. However, an examination of the participation ratio shows that in the correlated case, the PR is always greater than in the uncorrelated case. Thus, the correlated modes are less localized. To better understand the origin of this affect, we computed the transmission of a plane wave across a single magnetoinductive dimer, finding the transmission in closed form, as a function of the inductive coupling, and the frequency mismatch. It is found that this transmission can be unity for some wavevectors inside a region in coupling-mismatch space that is computed numerically. Thus, for plane waves whose wavevectors are close to resonance, there is a fraction of these modes whose localization length is of the size of the system.  On the dynamical front, computation of the mean square displacement shows that, while for the uncorrelated case there is saturation, in the correlated case there is finite propagation for small and medium disorder levels, and converging to smaller and smaller propagation at large disorder level. The correlation in the disorder also affected the transmission of plane waves across a finite disordered segment, where now the transmission decrease across the sample following a power-law decrease with sample's length instead of the typical exponential decrease. Finally,  
we compared the MSD for our system with  Dunlap's tight-binding RDM model, which also displays finite transport. While the MSD for the RDM model shows larger transport exponents at low disorder levels, at medium and high disorder strengths, our correlated model features larger exponent and thus greater mobility.  

We conclude that an array of SRRs with  uncorrelated disorder,  always displays Anderson localization of magnetic energy at any disorder strength. For correlated disorder at small and medium strengths, the SRR array  is capable of exhibiting a finite degree of mobility. At large disorder level, this mobility ceases and the system becomes Anderson-like.  These results could be useful for the design of efficient magnetic energy confinement devices, and the harvesting and transport of magnetic energy. We are current pursuing an extension of these studies to two-dimensional SRR systems. 

\acknowledgments

This work was supported by Fondecyt Grant 1200120.


\begin{thebibliography}{99}


\bibitem{SRR1}
T. J. Yen, W. J. Padilla, N. Fang, D. C. Vier, D. R. Smith, J. B. Pendry, D. N. Basov, and X. Zhang, Science
{\bf 303}, 1494 (2004).

\bibitem{SRR2}
N. Katsarakis, G. Constantinidis, A. Kostopoulos, R. S.
Penciu, T, F, Gundogdu, M. Kafesaki, E. N. Economou, Th. Koschny and C. M. Soukoulis, Opt. Lett. {\bf 30}, 1348 (2005).

\bibitem{SRR3}
M. I. Molina, N. Lazarides, and G. P. Tsironis, Phys. Rev. E {\bf 80}, 046605 (2009).
\bibitem{pendry}
J. B. Pendry, A. J. Holden, D. J. Robbins, and W. J. Stewart, IEEE Trans. Microwave Theory Tech. {\bf 47}, 2075 (1999). 

\bibitem{veselago}
V. G. Veselago, Sov. Phys. Usp. {\bf 10}, 509 (1968).

\bibitem{negative_refraction}
R. A. Shelby, D. R. Smith, S. Schultz, Science {\bf 292}, 77 (2001).

\bibitem{padilla}
D. Smith, W. Padilla, D. Vier, S. Nemat-Nasser, and S. Schultz,
Phys. Rev. Lett. {\bf 84}, 4184 (2000).



\bibitem{losses1}
L. Esaki, Phys. Rev. {\bf 109}, 603 (1958).

\bibitem{losses2}
T. Jiang, K. Chang, L.-M. Si, L. Ran, and H. Xin, Phys.
Rev. Lett. {\bf 107}, 205503 (2011).

\bibitem{anderson1} 
P. W. Anderson, Phys. Rev, 109, 1492 (1958).

\bibitem{anderson2} 
B. Kramer and A. MacKinnon, Rep. Prog. Phys. 56, 1496 (1993).

\bibitem{anderson3}
Elihu Abrahams, 50 Years of Anderson Localization (World Scientific Publishing, 2010).

\bibitem{correlated1} 
J. C. Flores, J. Phys. Cond. Matt. 1, 8471 (1989).

\bibitem{correlated2} 
D.H Dunlap. H. L. Wu. and P. W. Phillips, Phys. Rev. Lett. 65, 88 (1990).

\bibitem{correlated3} 
D. Giri, P.K. Datta, and K. Kundu, Phys. Rev. B 48, 14113 (1993).

\bibitem{szameit}
U. Naether, S. Stuetzer, R. Vicencio, M. I. Molina, A. T\"{u}nnermann, S. Nolte, T. Kottos, D. N Christodoulides and A. Szameit, New J. Phys. 15, 013045 (2013).

\bibitem{nmers}
D. Lopez and M. I. Molina, Phys. Rev. E. {\bf 93}, 032205 (2016).



\end{thebibliography}
\end{document}